# Superconductivity at 3.1 K in the orthorhombic ternary silicide ScRuSi

Bin-Bin Ruan[1], Xiao-Chuan Wang[1], Jia Yu[1], Bo-Jin Pan[1], Qing-Ge Mu[1], Tong Liu[1], Gen-Fu Chen[1, 2, 3], Zhi-An Ren[1, 2, 3] *

[1] Institute of Physics and Beijing National Laboratory for Condensed Matter Physics, Chinese Academy of Sciences, Beijing 100190, China

[2] Collaborative Innovation Center of Quantum Matter, Beijing 100190, China

[3] School of Physical Sciences, University of Chinese Academy of Sciences

* Email: renzhian@iphy.ac.cn

**Abstract:**

We report the synthesis, crystal structure, superconductivity and physical property characterizations of the ternary equiatomic compound ScRuSi. Polycrystalline samples of ScRuSi were prepared by an arc-melting method. The as-prepared samples were identified as the orthorhombic $Co_2P$-type $o$-ScRuSi by the powder X-ray diffraction analysis. Electrical resistivity measurement shows $o$-ScRuSi to be a metal which superconducts below a $T_c$ of 3.1 K, and the upper critical field $\mu_0H_{c2}(0)$ is estimated to be 0.87 T. The magnetization and specific heat measurements confirm the bulk type-II superconductivity in $o$-ScRuSi, with the specific heat jump within the BCS weak coupling limit. $o$-ScRuSi is the first $Co_2P$-type superconductor containing scandium. After annealing at 1273 K for a week, $o$-ScRuSi transforms into the hexagonal $Fe_2P$-type $h$-ScRuSi, and the latter is a Pauli-paramagnetic metal with no superconductivity observed above 1.8 K.

## 1. Introduction

The ternary equiatomic compounds $TT'X$ ($T, T'$ = transition metals; $X$ = Al, Ga, Si, Ge, Sn, P, As, Sb, Bi, *etc.*) have been intensively studied due to their various intriguing structural and magnetic properties [1]. For instance, Kondo effect [2], spin glass behavior [3] or large magnetoresistance [4] emerges when $T'$ is a 3d-metal. Superconductivity usually happens when $T'$ is a noble metal and $X$ stands for Al, P, As, Si or Ge [5-14], as well as in MoNiP [12] and a few compounds containing La or Bi [15, 16]. These superconductors were mainly classified into three structural types: (a) the hexagonal $Fe_2P$-type (or ZrNiAl-type) [10, 17]; (b) the orthorhombic $Co_2P$-type (or TiNiSi-type) [5, 9]; (c) the orthorhombic TiFeSi-type [8]. The $Co_2P$-type MoRuP marks the highest superconducting transition temperature ($T_c$) of 15.5 K among all these superconductors [13].

Many $TT'X$-type superconductors are formed when $T = Tr$ ($Tr$ = Ti, Zr or Hf), such as $Tr$RuP [17], $Tr$RhSi [9] and $Tr$IrSi [14]. For $T$ = rare-earth elements, only some La-containing compounds were discovered to date [15]. We noticed that Sc is close to $Tr$ in the atomic radius, and forms many similar crystal structures [18, 19]. Therefore, new superconductors could be anticipated in Sc-containing $TT'X$-type compounds. Actually, superconductivity has already been discovered in the $Fe_2P$-type ScIrP and ScRhP very recently [20, 21].

ScRuSi is an analog to ZrRuSi which is a superconductor with a moderate $T_c$ = 12 K [7]. It was first discovered as the hexagonal $Fe_2P$-type in 1982 [22]. However, Harmening *et al.* reported ScRuSi to crystallize in the orthorhombic $Co_2P$-type phase in an NMR (nuclear magnetic resonance) study in 2011 [18], in sharp contrast to the former one. In fact, ScRuSi can crystallize in either the $Co_2P$-type or the $Fe_2P$-type phase by our investigation, which is hereafter referred to as $o$-ScRuSi or $h$-ScRuSi, respectively. $h$-ScRuSi has a layered structure similar to that of superconducting hexagonal ZrRuSi, but it was predicted to be less likely to superconduct due to the extraordinarily long distance of the intra-layer Ru atoms [23]. Experimentally, neither of the two polymorphs of ScRuSi has been systematically examined for their low temperature physical properties.

In this paper, we studied the ternary ScRuSi compound and successfully prepared both of the *o*-ScRuSi and *h*-ScRuSi samples. Superconductivity at $T_c$ = 3.1 K was observed in *o*-ScRuSi, while *h*-ScRuSi did not superconduct down to 1.8 K. The crystal structure and low temperature physical properties for both polymorphs were characterized. We newly noticed that the superconductivity of *o*-ScRuSi was also reported in a brief meeting abstract in 2011 without more detailed characterizations [24].

## 2. Experimental

Polycrystalline samples of ScRuSi were synthesized by an arc-melting method. The starting materials were ruthenium powder, silicon powder and scandium foil, all purer than 99.9%. In order to avoid shattering during the melting reaction, the Sc foils were pre-melted into buttons under purified argon. Powders of elemental Ru and Si were mixed in a molar ratio of 1:1, pelletized and then mixed with a stoichiometric amount of Sc buttons. The mixtures were arc-melted under an argon atmosphere for at least ten times with intermediate turnovers to ensure the sample homogeneity. The weight loss after arc-melting was always less than 1%. Some of the samples were directly measured, while the others were wrapped with tantalum foils, sealed into evacuated quartz tubes, and annealed at 1273 K for a week. All the samples showed metallic luster and were stable in air, whether annealed or not. Superconductivity was observed only in the as-prepared samples, and was absent above 1.8 K in the annealed ones.

Powder X-ray diffraction (PXRD) was measured for all samples on a PAN-analytical X-ray diffractometer with Cu-K$_\alpha$ radiation at room temperature for crystal structure analysis. Rietveld refinement of the PXRD data was carried out using the EXPGUI interface for the GSAS package [25].

The electrical resistivity and thermal properties were measured on a physical property measurement system (PPMS, Quantum Design). The resistivity was measured by a standard four probe method. The magnetization was measured by a magnetic property measurement system (MPMS, Quantum Design). Notice that the structural and corresponding transport properties were measured on the same specimen.

## 3. Results and discussion

In Figure 1 we show the PXRD data of the as-prepared polycrystalline ScRuSi sample collected from 20° to 90° at room temperature, as well as its Rietveld analysis. Except for tiny amount of $Ru_2Si$ impurity (marked by the hash), all the diffraction peaks were indexed by the orthorhombic space group *Pnma* (No. 62), revealing the sample to be *o*-ScRuSi. The structure refinement was performed with the atom occupancies fixed to 1, and it ended up with $\chi^2$ = 2.89, $R_{wp}$ = 2.87%, giving the lattice parameters: $a$ = 6.622(2) Å, $b$ = 4.100(1) Å, and $c$ = 7.044(6) Å. The crystallographic data from refinement are summarized in Table I. These results agree well with those in the literature [18]. The crystal structure is shown as the inset of Figure 1, where Ru atoms form a three-dimensional network with Si atoms in a distorted tetrahedral environment, and Sc atoms are filled in the cavities. The chemical composition was also confirmed by the energy-dispersive spectroscopy (EDS) method, with the atomic ratio of Sc:Ru:Si close to 1:1:1 within the instrumental error.

Figure 2(a) shows the temperature dependence of electrical resistivity from 1.8 K to 300 K under zero magnetic field. The resistivity is 1.04 mΩ cm at 300 K and descends with a decreasing temperature, showing a metallic behavior in the whole temperature range. A sharp superconducting transition is observed at the onset temperature of $T_c^{onset}$ = 3.1 K and zero resistivity occurs below 2.5 K, giving a superconducting transition width of 0.6 K. The relatively broad transition is due to the sample inhomogeneity, which comes from the arc-melting process without annealing. Magnetic fields up to 0.5 T were applied to investigate the field dependence of superconductivity, and the results are shown in Figure 2(b). No significant magnetoresistance is observed in the normal state. $T_c^{onset}$ decreases gradually with increasing magnetic field, indicating the suppression of superconductivity by the field. The mid-point of the superconducting transition is defined as $T_c^{mid.}$, whose dependence with magnetic field is depicted in the inset of Figure 2(a). The upper critical field $\mu_0 H_{c2}(0)$ is estimated to be 0.87 T by the Ginzburg-Landau formula: $H_{c2}(T) = H_{c2}(0)(1 - t^2)/(1 + t^2)$, where $t$ is the normalized temperature $T/T_c^{mid.}$. Thus the Ginzburg-Landau coherence length $\xi(0)$ is calculated to be ~ 168 Å. The relatively large $\xi(0)$ suggests the sample to be quite clean.

The temperature dependence of DC magnetic susceptibility was measured under a magnetic field of 5 Oe from 1.8 K to 300 K in both zero-field-cooling (ZFC) and field-cooling (FC) process. The results are shown in Figure 3(a) with an enlarged view in the inset. The diamagnetic superconducting transitions are detected below 2.7 K in both the ZFC and FC runs. The superconducting shielding volume fraction is estimated to be 94% at 1.8 K from the ZFC data, suggesting bulk superconductivity in $o$-ScRuSi. The isothermal magnetization loop was measured at 2 K, and the results are depicted in the inset of Figure 3(b). The features of the loop reveal the sample to be a typical type-II superconductor. Details under low fields are shown in Figure 3(b). The red dash line is a linear fit of the initial data. The lower critical field at 2 K ($\mu_0 H_{c1}(2\ K)$) is thus determined to be 1.4 mT from the deviation between the measured data and the linear fitting.

To further investigate the bulk superconductivity in $o$-ScRuSi, the low temperature specific heat was measured under the fields of $\mu_0 H = 0$ and 2 T from 1.8 K to 6 K. The results are demonstrated as $C_p/T$ versus $T^2$ in Figure 4(a). A large specific heat jump of the zero field curve below 3.0 K confirms bulk superconductivity in the sample. The normal state specific heat is well fitted with the formula $C_p/T = \gamma + \beta T^2$, giving $\gamma = $ 12.08 mJ mol$^{-1}$ K$^{-2}$ and $\beta = 0.10$ mJ mol$^{-1}$ K$^{-4}$. Thus the Debye temperature ($\Theta_D$) is estimated to be about 388 K from the simple Debye model. These parameters are close to that of the isostructural superconductor ZrIrSi [14]. To study the superconducting transition, the electronic contribution to specific heat ($C_e$) is plotted in Figure 4(b). The transition is relatively broad (from 2.0 K to 2.9 K), possibly related to the inhomogeneity. The superconducting transition temperature from specific heat ($T_c^{heat}$) is determined to be 2.35 K from the jump. The specific heat of electrons is tentatively fitted below $T_c^{heat}$ with the α model [26], as depicted in Figure 4(b). The specific heat jump at $T_c^{heat}$ ($\Delta C_p(T_c^{heat})$) is estimated to be about 34.88 mJ mol$^{-1}$ K$^{-1}$, thus $\Delta C_p(T_c^{heat})/\gamma T_c^{heat}$ equals to 1.23, which is within the value of 1.43 for the BCS weak coupling limit.

As for the annealed sample, the PXRD pattern is shown in Figure 5, indicating that the sample transformed into a hexagonal Fe$_2$P-type $h$-ScRuSi completely from the as-prepared $o$-ScRuSi, with minor impurities of Ru$_2$Si detected. This kind of structural phase transition also happens in ScCuSi, which transforms from an orthorhombic phase into a low-temperature hexagonal one after annealing at 1073 K [27]. $h$-ScRuSi has a space group of $P$-62$m$ (No. 189) with refined lattice parameters: $a = b = 6.878(6)$ Å and $c = 3.375(4)$ Å. The crystal structure is depicted as the inset of Figure 5. $h$-ScRuSi consists of alternating Ru$_2$Si and Sc$_2$Si layers, while Ru atoms maintain a distorted tetrahedral coordination with Si atoms. The temperature dependences of resistivity and magnetic susceptibility were measured for the $h$-ScRuSi samples above 1.8 K, with no superconductivity discovered, as shown in Figure 6. The absence of superconductivity in $h$-ScRuSi is consistent with a previous half-empirical prediction [23]. The magnetic susceptibility was fitted by the Curie-Weiss formula: $\chi(T) = \chi_0 + C/(T + \theta)$ with $\chi_0 = 4.37\times10^{-4}$ emu mol$^{-1}$ Oe$^{-1}$, $C = 1.37\times10^{-3}$ emu K mol$^{-1}$ Oe$^{-1}$ and $\theta = 3.36$ K. The local moment was estimated to be ~ $0.10\mu_B$ per mole where $\mu_B$ is the Bohr magneton. These parameters are comparable with those in typical ruthenium alloys such as LaRu$_3$Si$_2$ [28], suggesting that $h$-ScRuSi is a Pauli-paramagnetic metal.

## 4. Conclusions

In summary, we have synthesized samples of ScRuSi by an arc-melting method. Their transport properties as well as the structural phase transition are reported. Physical property measurements of $o$-ScRuSi reveal it to be a bulk type-II superconductor with a $T_c$ = 3.1 K, serving as a new member of the $Co_2P$-type superconductors. After annealing at 1273 K for a week, *o*-ScRuSi transformed into *h*-ScRuSi, which turned out to be a Pauli-paramagnetic metal without superconductivity above 1.8 K.

**Acknowledgments**

The authors are grateful for the financial supports from the National Natural Science Foundation of China (No. 11474339), the National Basic Research Program of China (973 Program, No. 2010CB923000 and 2011CBA00100) and the Strategic Priority Research Program of the Chinese Academy of Sciences (No. XDB07020100).

**Table 1.** Structural parameters and atomic coordinates of *o*-ScRuSi.

| Chemical formula | | | ScRuSi | | | |
|---|---|---|---|---|---|---|
| **Crystal system** | | | Orthorhombic | | | |
| **Space group** | | | *Pnma* (No. 62) | | | |
| **Z** | | | 4 | | | |
| **Theoretical density (g/cm$^3$)** | | | 6.0481(9) | | | |
| **Lattice parameters** | | | | | | |
| *a* (Å) | | | 6.6222(7) | | | |
| *b* (Å) | | | 4.1001(4) | | | |
| *c* (Å) | | | 7.0446(9) | | | |
| **Cell volume *V* (Å$^3$)** | | | 191.27(8) | | | |
| Atom | Site | x | y | z | Occupancy | $B_{iso}$ (Å$^2$) |
| Sc | 4*c* | -0.0012(1) | 0.25 | 0.6844(1) | 1 (fixed) | 0.1373(9) |
| Ru | 4*c* | 0.1592(5) | 0.25 | 0.0592(2) | 1 (fixed) | 0.4121(7) |
| Si | 4*c* | 0.2914(8) | 0.25 | 0.3900(4) | 1 (fixed) | 0.9246(2) |

**Figure captions:**

**Figure 1.** Powder X-ray diffraction pattern of the *o*-ScRuSi sample (red cross) and the Rietveld refinement (black curve). Tiny Ru$_2$Si impurity is marked by the hash. The inset shows the crystal structure of *o*-ScRuSi.

**Figure 2.** (a) Temperature dependence of resistivity for the *o*-ScRuSi sample from 1.8 K to 300 K. The inset shows the upper critical field $\mu_0 H_{c2}$ determined by the mid-points of the superconducting transitions ($T_c^{mid.}$) under various magnetic fields. The solid curve is the Ginzburg-Landau fitting of $\mu_0 H_{c2}$. (b) Temperature dependence of resistivity for the *o*-ScRuSi sample from 1.8 K to 4 K under various magnetic fields.

**Figure 3.** (a) Temperature dependence of magnetic susceptibility under a field of 5 Oe for the *o*-ScRuSi sample in the zero field cooling (ZFC) and the field cooling (FC) process. The inset shows details around the superconducting transition. (b) Field dependence of magnetization at 2 K. The inset shows the superconducting loop of the *o*-ScRuSi sample.

**Figure 4.** (a) Low temperature specific heat under magnetic fields of $\mu_0 H = 0$ T and 2 T

from 1.8 K to 6 K. (b) The electronic specific heat data under zero magnetic field. The solid curve shows the fitting by the $\alpha$ model.

**Figure 5.** Powder X-ray diffraction patterns of the *h*-ScRuSi sample. The blue bars indicate the Bragg peaks of the main phase. Minor $Ru_2Si$ impurity is marked by the hash. The inset shows the crystal structure of *h*-ScRuSi.

**Figure 6.** (a) Temperature dependence of resistivity for the *h*-ScRuSi sample from 1.8 K to 300 K. (b) Temperature dependence of magnetic susceptibility under a field of 0.5 T for the *h*-ScRuSi sample. The Curie-Weiss fit $\chi = \chi_0 + C/(T + \theta)$ is displayed as the red curve.

**Figure 1.**

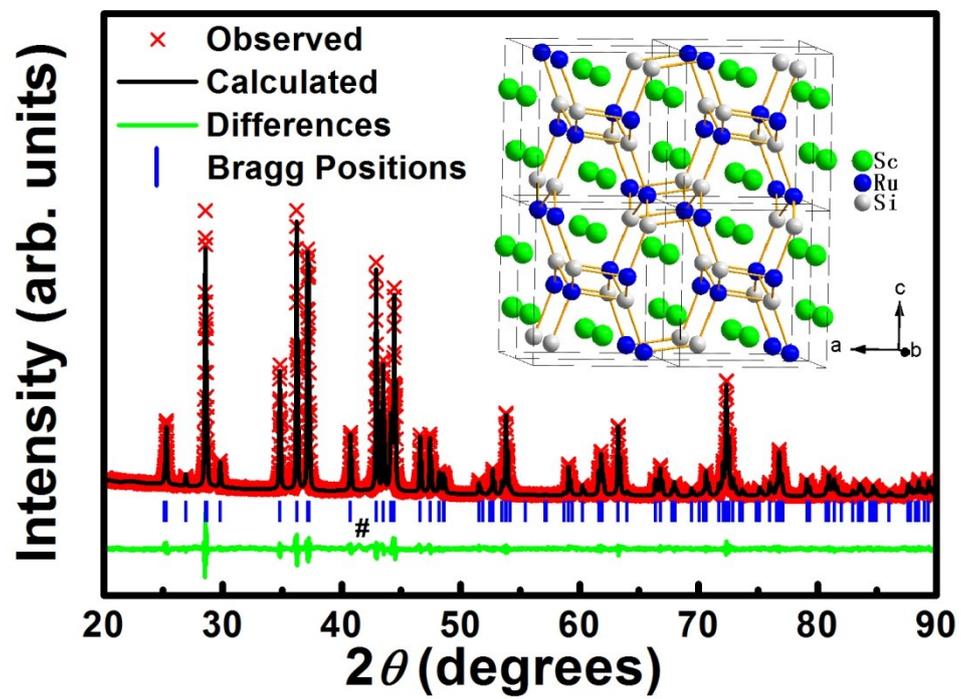

**Figure 2.**

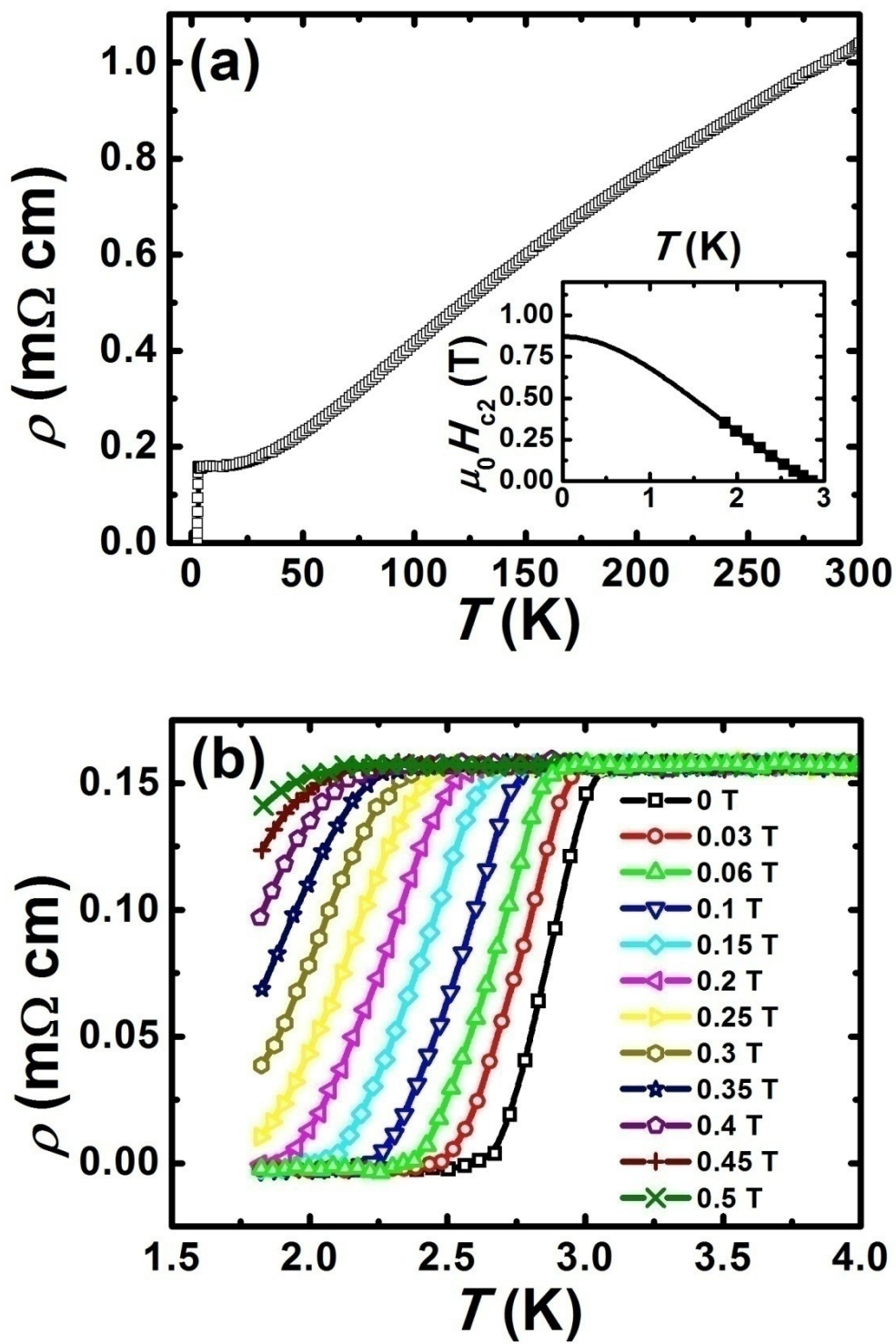

**Figure 3.**

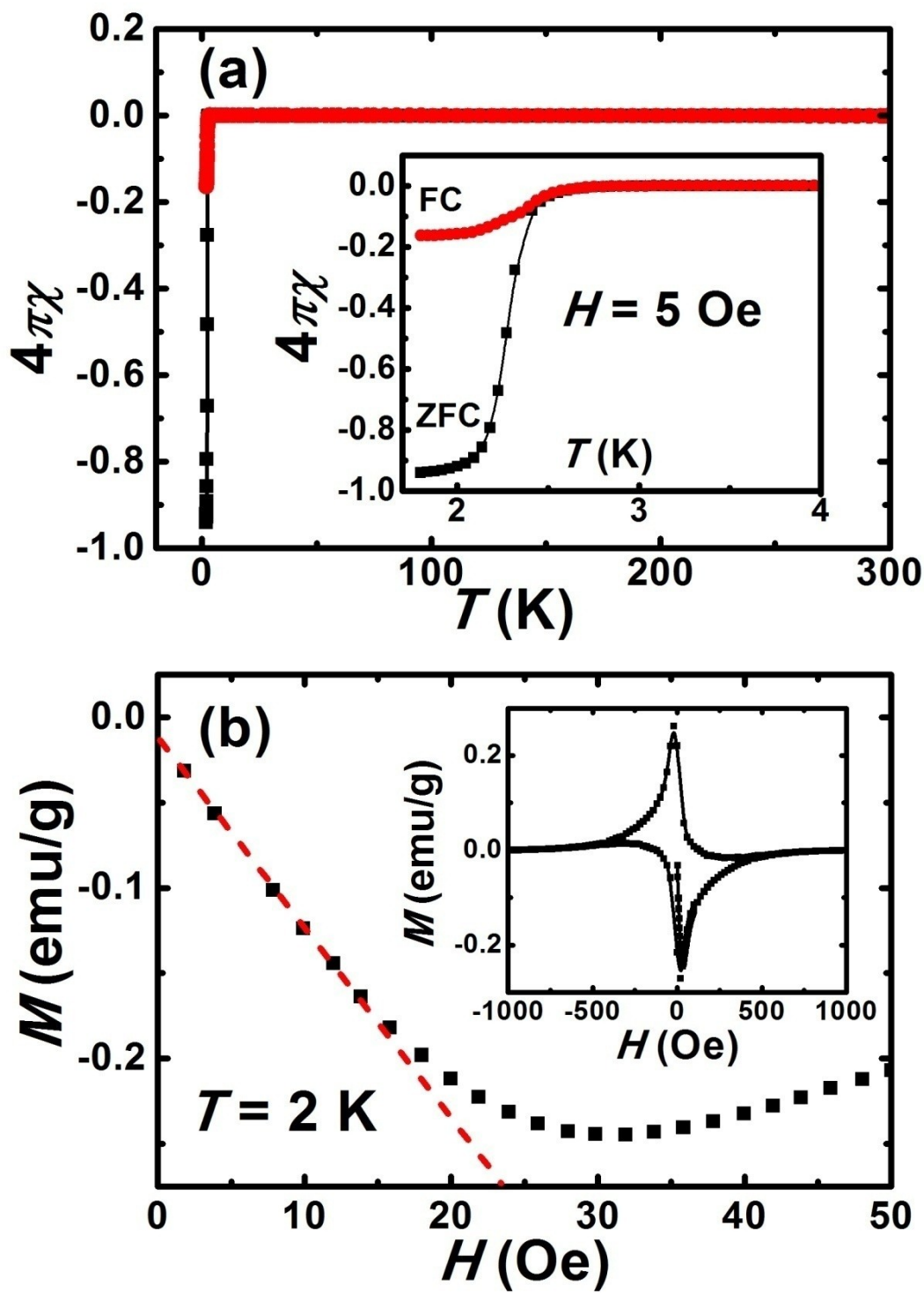

**Figure 4.**

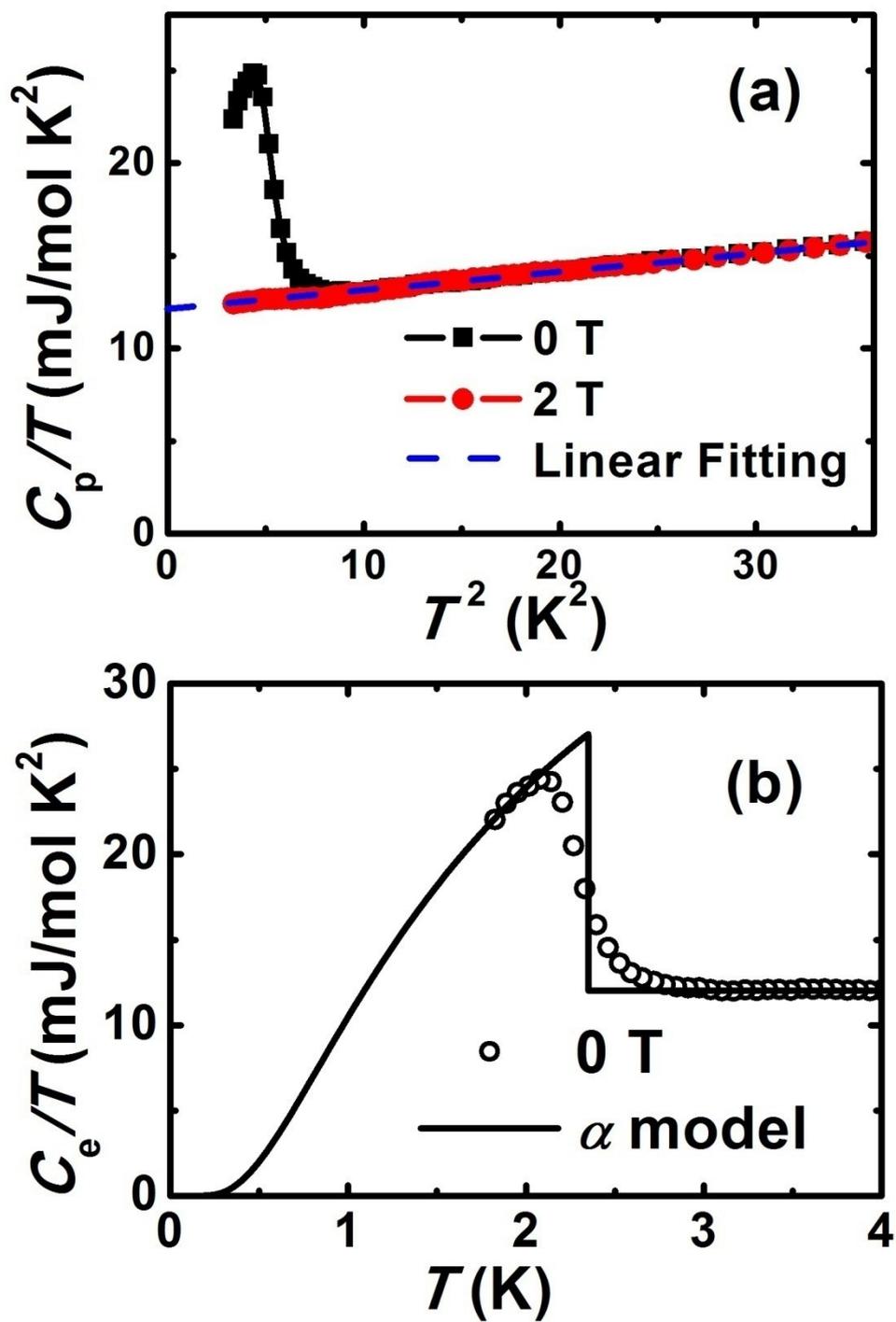

**Figure 5.**

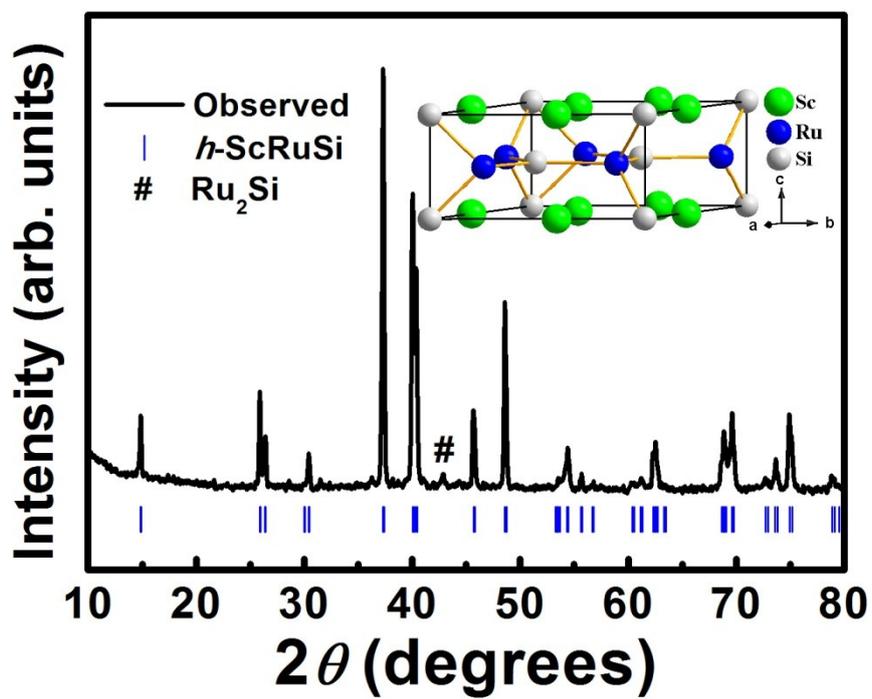

**Figure 6.**

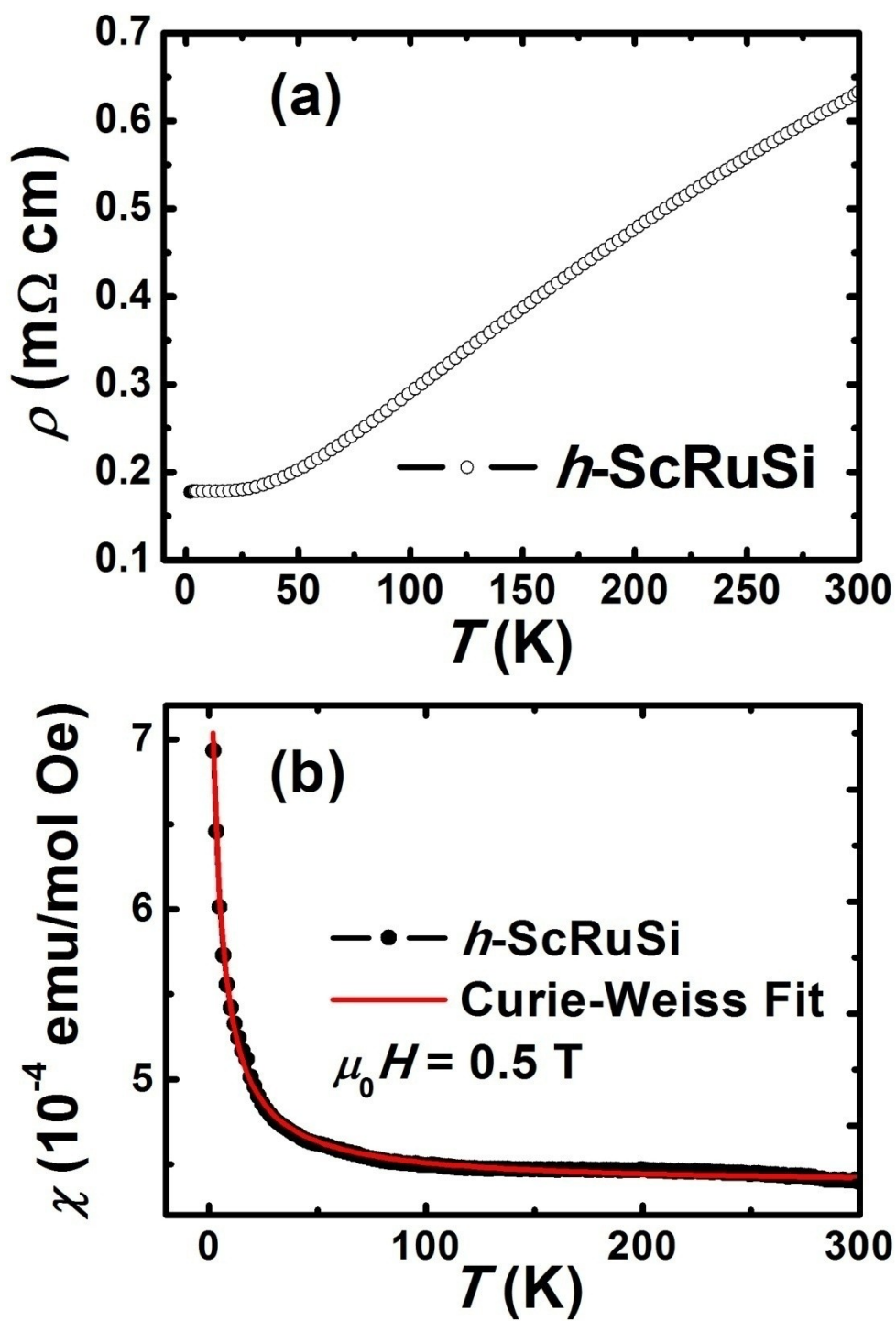